\documentclass[twocolumn,showpacs,preprintnumbers,amsmath,amssymb,superscriptaddress]{revtex4}
\pdfoutput=1
\usepackage{graphicx}
\usepackage{dcolumn}
\usepackage{bm}

\begin{document}

\title{
Quantum Griffiths Phase in the weak itinerant ferromagnetic alloy Ni$_{1-x}$V$_x$}
\author{Sara Ubaid-Kassis}
\affiliation{Physics Department, Kent State University, Kent OH 44242}
\author{Thomas Vojta}
\affiliation{Department of Physics, Missouri University of Science and Technology, Rolla, MO 65409}
\author{Almut Schroeder}%
\affiliation{Physics Department, Kent State University, Kent OH 44242}

\date{\today}

\begin{abstract}
We present magnetization ($M$) data of the $d$-metal alloy Ni$_{1-x}$V$_x$ at vanadium
concentrations close to $x_c \approx 11.4\%$ where the onset of long-range ferromagnetic
(FM) order is suppressed to zero temperature. Above $x_c$, the temperature ($T$) and
magnetic field ($H$) dependencies of the magnetization are best described by simple
nonuniversal power laws. The exponents of $M/H \sim T^{-\gamma}$ and   $M \sim H^\alpha$
are related by $1-\gamma=\alpha$ for wide temperature ($10K < T \leq 300K$) and  field
($H \leq 5T$) ranges. $\gamma$ is strongly $x$ dependent, decreasing from 1 at $x\approx
x_c$ to $\gamma < 0.1$ for x=15$\%$. This behavior is not compatible with either
classical or quantum critical behavior in a clean 3D FM. Instead it closely follows the
predictions for a quantum Griffiths phase associated with a quantum phase transition in a
disordered metal. Deviations at the lowest temperatures hint at a freezing of large
clusters and the onset of a cluster glass phase, presumably due to RKKY interactions in
this alloy.

\end{abstract}

\maketitle


Magnetic phase transitions in metals continue to offer challenges to theory and
experiment. In recent years, the focus has shifted from thermal transitions (such as the
onset of ferromagnetism in nickel at a temperature of 630K \cite{weiss26})
to quantum phase transitions (QPTs) \cite{sachdev99} that occur at zero temperature when
a parameter such as pressure or chemical composition is varied. Spin fluctuations
associated with continuous QPTs or quantum critical points (QCPs) are believed to be
responsible for a variety of exotic phenomena including deviations from the
fundamental Fermi liquid behavior of normal metals.

Ferromagnetic and antiferromagnetic QCPs have been observed in transition metal
alloys and heavy-fermion compounds (see Ref.\ \cite{hvl07} for a review).
Quantum critical behavior is signified by singularities in thermodynamic and transport
properties. According to the standard theory of ferromagnetic quantum criticality in 3D
metals \cite{HertzMillis}, specific heat $C$, magnetic susceptibility $\chi$ and
electrical resistivity $\rho$ should behave as $\chi \sim T^{-4/3}$,  $C/T \sim \log(T)$
and  $\rho(T) \sim T^{5/3}$ when approaching the QCP at low temperatures $T$. This was
observed in Ni$_{x}$Pd$_{1-x}$ with $x=0.025$, at least over a limited temperature regime
\cite{nicklas99nipd}. However, most ``clean" weak ferromagnets like MnSi
\cite{pfleiderer97}, ZrZn$_2$ \cite{uhlarz04} or Ni$_3$Al \cite{niklowitz05ni3al} show
deviations from the above predictions. The QPT becomes first order \cite{BKV} and is
often accompanied by the appearance of novel phases.

Many ferromagnetic binary alloys such as Ni$_{1-x}$Cu$_x$ or Ni$_{1-x}$V$_x$
in which $T_c$ can be tuned by chemical substitution $x$ show a still more complicated
behavior, even in the paramagnetic phase.
In early investigations \cite{amamou75rho}, the existence of large magnetic clusters with
giant local moments was proposed to describe the magnetization $M$ data of these inhomogeneous
systems.

Recent theories address the impact of disorder on QPTs more systematically
(for a review, see Ref.\ \cite{vojta06}). We are
interested in the case of metallic (itinerant) Heisenberg magnets. For these systems a
strong-disorder renormalization group  \cite{vojta0709} predicts an
exotic infinite-randomness QCP, accompanied by quantum Griffiths singularities
\cite{vojta05}. At such a QCP, thermodynamic observables are expected
to be singular not just at criticality but in a finite region around the QCP called the
Griffiths phase. This is caused by rare spatial regions that are locally in the
magnetic phase while the bulk is still nonmagnetic. The probability
$w$ of finding such regions is exponentially small in their volume $V$, $w \sim
\exp(-bV)$ with $b$ a constant that depends on the disorder strength.
Importantly, the characteristic energy scale $\epsilon$ of a locally ordered region
also depends exponentially on its volume, $\epsilon \sim \exp(-cV)$. Combining these two
exponentials yields an energy spectrum $P(\epsilon) \sim \epsilon^{\lambda-1}$.
The nonuniversal Griffiths exponent $\lambda=b/c$ takes the value 0 at the quantum critical point
and increases with distance from criticality. This power-law spectrum gives
rise to power-law quantum Griffiths singularities of many observables, including
specific heat $C\sim T^\lambda$, susceptibility $\chi \sim T^{\lambda-1}$ and
the zero temperature magnetization-field curve, $M\sim H^\lambda$.

Quantum Griffiths singularities have attracted a lot of attention,
but clearcut experimental verifications have been slow to arrive. Many heavy fermion compounds
display anomalous power-laws in $C(T)$ and $\chi(T)$  \cite{stewart01};
and quantum Griffiths behavior was suggested as an explanation \cite{castro00}.
However, in most of these systems, a systematic variation of the exponents in accordance
with theory could not be observed. Only recently, a partial confirmation could be
found at the ferromagnetic QPT of CePd$_{1-x}$Rh$_{x}$ \cite{westerkamp09CePd}.
It must also be noted that the interpretation of experiments in heavy-fermion compounds
suffers from additional complications due to the Kondo effect which plays a crucial role
for the magnetic properties. It is thus highly desirable to observe quantum
Griffiths singularities in a simpler system.

In this Letter we therefore study the transition metal alloy  Ni$_{1-x}$V$_x$ as an example of an
itinerant ferromagnet in which $T_c$ can be tuned to zero by chemical substitution while
introducing strong disorder. We show that magnetization and susceptibility close to the
critical vanadium concentration indeed follow the quantum Griffiths scenario over a wide
temperature and magnetic field range.



Polycrystalline spherical Ni$_{1-x}$V$_x$ samples with $x=9-15\%$ were prepared by arc
melting from high purity elements (Ni 99.995$\%$,V 99.8$\%$) and annealed at 900 - 1050$^o$C. X-ray diffraction
confirmed a single phase fcc-structure with lattice constant $a=(0.352+0.023 x)nm$.
Magnetization measurements were performed in a Quantum Design SQUID magnetometer from
1.8K - 300K and magnetic fields up to 5T. The ac-susceptibility was measured in a pick up
coil in a dilution refrigerator down to 0.05K and calibrated through the overlap with the
magnetometer data. All data shown are demagnetized.

It is known that $T_c$ of Ni$_{1-x}$V$_x$ is rapidly reduced with increasing
V-concentration $x$ \cite{boelling68}. As explained by Friedel \cite{friedel58},
Ni$_{1-x}$V$_x$ resides on a side branch of the Slater-Pauling curve: a V impurity (with
5 fewer electrons than Ni) creates a localized charge and a spin reduction on the
neighboring Ni-sites. This reduces the average spin moment by
$5\mu_B/V$ from $0.6\mu_B/$Ni leading to a critical concentration of about 12$\%$. It
also creates large defects yielding an inhomogeneous magnetization density which makes
Ni$_{1-x}$V$_x$ an ideal compound to study a QPT with significant ``disorder".

\begin{figure}
\includegraphics{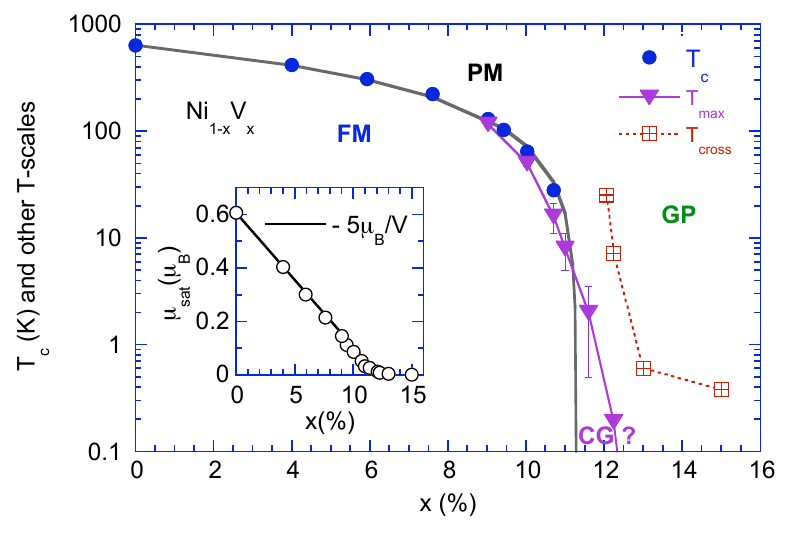}
\caption{\label{fig:pd} Temperature-concentration phase diagram of Ni$_{1-x}$V$_x$
showing ferromagnetic (FM), paramagnetic (PM), quantum Griffiths (GP), and cluster glass
(CG) phases. Circles mark $T_c$ found via the Arrott analysis. The grey line is an
extrapolation of $T_c(x)$. Also shown are $T_{max}$ defined by low-field maxima in $\chi(T)$
and $T_{cross}$ below which frozen clusters dominate $\chi(T)$, leading to
superparamagnetism.
Inset: saturation magnetization $\mu_{sat}$ vs $x$ (determined as $M$($T<5$K,
$H>1$T)).
Data from Ref.\ \cite{boelling68} for $x<11\%$ are included.}
\end{figure}

The phase diagram
resulting from our
measurements is shown in Fig.\ \ref{fig:pd}.
To get an overview,
we first perform a standard Arrott analysis. Samples with  $x \le 11\%$
show mean-field behavior, i.e., parallel isotherms of the form $H/M=a+bM^2$ as is common
for itinerant magnets. (Mean-field behavior is expected outside the actual critical region;
if the standard theory \cite{HertzMillis} applied, it would describe the entire
transition up to log.\ corrections.) $T_c$ is extracted from the Arrott plots via
the mean-field $T$-dependencies of magnetization and susceptibility.

For $x>11\%$, clear deviations occur from mean-field behavior (linear Arrott plots) \cite{ubaid08};
and the determination of $T_c$ becomes sensitive to model assumptions. In addition to Arrott
plots, we analyze the differential
susceptibility $\chi(T) = dM(T)/dH$. It exhibits a field-dependent maximum at $T_{max}(H)$,
indicating spin ordering or freezing.
Figure \ref{fig:pd} shows $T_{max}(H \rightarrow 0)$ estimated by a linear
extrapolation of the data taken at 0.5T to 0.1T. $T_{max}(H \to 0)$ is somewhat lower than $T_c$
derived from the high field mean field analysis.
We note that the magnetization
for $x>11\%$ and \emph{higher} fields $H>0.5T$ can be described by parallel isotherms in a modified
Arrott plot \cite{arrott67} ($(H/M)^{(1/\gamma)}=a+bM^{(1/\beta)}$) with exponents $\beta=0.5$
and an $x$-dependent $\gamma(x)<1$ \cite{ubaid08}. However, this yields a finite
$T_c$ well above $T_{max}$ for all $x \le 15\%$.

\begin{figure}
\includegraphics{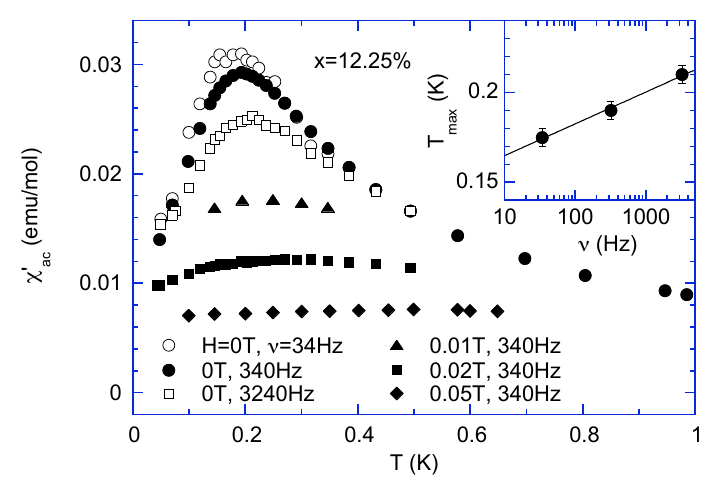}
\caption{\label{fig:chi_ac} ac-susceptibility $\chi_{ac}$ of the $x=12.25\%$ sample for several
dc magnetic fields $H$ and frequencies $\nu$ of the (in phase) ac-magnetic field ($H_{ac} \approx 0.1$G) vs
temperature $T$. Inset shows the frequency dependence of the maximum in $\chi(T)$ $T_{max}$.
The line follows $dT_{max}/d(\log\nu)=0.018$K$/dec(\nu)$.}
\end{figure}

Within our error bars, $T_{max}(H \rightarrow 0)$ is definitely nonzero for $x \le 11\%$ but
zero for $x \ge13\%$. To understand the behavior at intermediate concentrations,
we measure the ac-susceptibility $\chi_{ac}$ in a small ac-field $H_{ac}\approx 0.1G$ down to 50mK
as shown in Fig.\ \ref{fig:chi_ac} for the sample with $x=12.25\%$.
A maximum in $\chi_{ac}(T)$ at $T_{max}=0.19K$ marks spin freezing in $0mT$ at a low frequency of $\nu=340$Hz.
It is rapidly suppressed in small dc-fields and shifted to higher $T$. $T_{max}$ is dependent on $\nu$
like in a spin glass, signifying irreversibility in this system. At higher $T$, no significant
hysteresis in $M(H)$ was found for all samples with a remanent field larger than the rest field of
the magnet of the order of 10G.

We emphasize that deviations from linear Arrott plots and the sensitivity of $T_c$
towards the extrapolation procedure already point to an unconventional QPT. Moreover,
the $x$-dependence of $T_{max}(H\rightarrow0)$ in the accessible temperature region is better
described by an exponential rather than a power law, making the determination of
the  $x_c$ from finite-temperature data difficult.

We now turn to the paramagnetic phase above the critical concentration $x_c \approx
11\%$. In the past, the susceptibility at higher temperatures ($T > 40$ K) has been
described \cite{amamou75rho} by a Curie-Weiss law $\chi = C/
(T-\theta)+\chi_{orb}$,
but this only works if the orbital contribution $\chi_{orb}$ is allowed to vary by a
factor of 3 with concentration $x$. In our analysis, we keep the orbital susceptibility
$x$-independent at $\chi_{orb}=6 \times 10^{-5}$emu/mol which is the best fit parameter
for $x<x_c$ and lies within the reported estimates \cite{shimizu64}.

\begin{figure}
\includegraphics{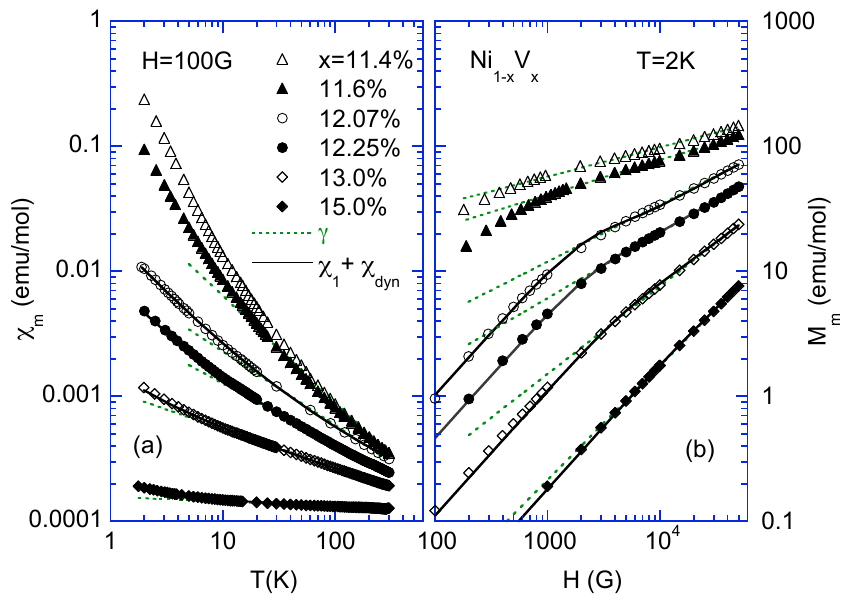}
\caption{\label{fig:M/H} (a) Low-field susceptibility $\chi_m=M/H-\chi_{orb}$ of Ni$_{1-x}$V$_x$
vs temperature $T$ and (b) low-temperature magnetization $M_m = M-\chi_{orb}H$ vs magnetic field
$H$  ($x=11 - 15\%$). Dotted lines indicate power laws for $T>10$ K and $H>3000$ G in (a) and
(b), respectively. Solid lines (shown for $x>12\%$) represent fits to the model (\ref{eq:chi1})
with frozen clusters.}
\end{figure}

\begin{figure}
\includegraphics{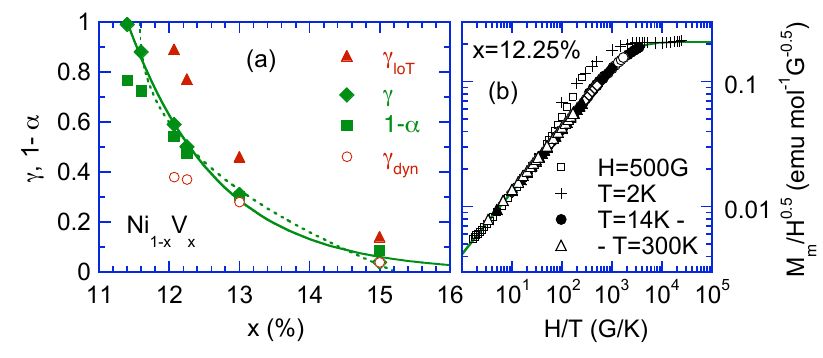}
\caption{\label{fig:exponents} (a) Exponents $\gamma, 1-\alpha$ obtained from $\chi_m \sim T^\gamma$ and
$M_m \sim H^\alpha$ vs V-concentration $x$. The plot also shows $\gamma_{loT}$ derived from
$\chi_m(T)$ below 10\,K and $\gamma_{dyn}$ extracted from
the fit to (\ref{eq:chi1}).
The $x$-dependence of $\gamma$ can be fitted 
to a power law, $(1-\gamma)
\sim (x-x_c)^{\nu\psi}$ (dotted line), as expected according to \cite{vojta0709}
(with $\nu\psi \approx 0.42$) or to an exponential,
$-\ln \gamma \sim (x-x_c)$ (solid line).
(b) $H/T$ scaling plot for $x=12.25\%$ showing  data at several const. $T$ from 2K - 300K and  $H=500$G. The line is a fit to the phenomenological
form discussed below (\ref{eq:scaling}). }
\end{figure}

The resulting low-field ($H=100$ G) spin susceptibility $\chi_m=M/H-\chi_{orb}$
is shown in Fig.\ \ref{fig:M/H}a for samples with $x=11 - 15\%$.
At temperatures above 10 K, simple power laws describes the data well.
We parametrize the power law by
$\chi_{m}(T) = A (k_B T)^{-\gamma}$. The exponent $\gamma$ is determined from fits between 30K and 300K
(300K may seem a very high temperature for analyzing a QPT, but it is still well below
the Curie point of Nickel at 630K; this high bare temperature scale is another advantage of our material.)
Figure \ref{fig:exponents}a
shows that $\gamma$ varies from about 1 for $x=11.4\%$  towards $0.03$ for $x=15\%$.
We also analyze the magnetization-field curve. Figure  \ref{fig:M/H}b shows
$M_m=M-\chi_{orb} H$ as a function of field  at the lowest $T=2K$
for samples with $x>11\%$. For fields above $H>3000G \approx k_B T/10\mu_B $, $M(H)$ follows a power law
with an exponent $\alpha$. Its value (shown in Fig.\ \ref{fig:exponents}a) matches the susceptibility
exponent, $1-\alpha= \gamma$  for $x>12\%$.

The results for $\chi_m(T)$ and $M_m(H)$ are in excellent agreement with the predictions
for a quantum Griffiths phase outlined in the introduction if we identify the Griffiths
exponent via $\lambda=\alpha = 1-\gamma$. The critical concentration $x_c$ can now be
determined from the condition $\lambda=0$ which gives $x_c \approx 11.4\%$ using the
susceptibility data. Right at criticality, the theory \cite{vojta0709} predicts extra
logarithmic corrections (which are notoriously hard to verify) to the power laws.
Our fits did not noticeably improve by including logarithmic terms.

To combine the temperature and field dependencies (at fixed $x$) of the magnetization we suggest
the scaling form
\begin{equation}
M = H^\alpha \, Y(\mu H/k_B T)
\label{eq:scaling}
\end{equation}
where $Y$ is the scaling function and $\mu$ is a scaling moment. The corresponding
scaling plot for $x=12.25\%$ is shown in Fig.\ \ref{fig:exponents}b. All $M(H,T)$ data
for temperatures above 14 K collapse, confirming $H/T$ scaling. The scaling function $Y$
is well approximated by the phenomenological form $Y(z)=A'/(1+z^{-2})^{\gamma/2}$ where
$A'=A/\mu^{\gamma}$ is a constant. We have produced similar scaling plots for the other
concentrations. The resulting exponent $\gamma$ matches that obtained by a direct fit of
$\chi(T)$ for all $x$ between 11.4\% and 15\%. The scaling moment $\mu$ increases from
$1\mu_B$ at $x=15\%$ to $12\mu_B$ at $x=11.4\%$ demonstrating the growth of the typical
cluster size  with $x \to x_c$ \cite{mag_footnote}. An analogous scaling form was used to describe the $H/T$
scaling in heavy fermions  \cite{schroder00}. It also  gives the correct exponent for the
nonlinear susceptibility ($\chi_3=d(M/H)/d(H^2) \sim T^{\alpha-3}$ for $T\gg H$).

Having established the quantum Griffiths phase, we now turn to its limits.
At $T<10$K, deviations are observed from the ``Griffiths'' power laws.
For instance, $\gamma$ increases by about $50\%$ (see Fig.\ \ref{fig:exponents}). This exponent
does not match $1-\alpha$. We thus believe that the behavior
below 10K deviates from the quantum Griffiths scenario either due to the Vanadium distribution
not being perfectly statistical or because the rare regions are
not independent.

It was shown \cite{dobro05} that RKKY interactions between the rare regions
lead to a dynamical freezing of the largest clusters
at low $T$ and to the formation of a cluster glass at even lower $T$.
To explore this possibility, we model the zero-field susceptibility
\begin{equation}
\chi_m(T) = \chi_1+\chi_{dyn}=  {A_1}/{(k_B T)} +  {A}/{(k_B T)^{\gamma_{dyn}}}
\label{eq:chi1}
\end{equation}
as the sum of a Curie term $\chi_1$ (describing the frozen clusters) and
a Griffiths term with an exponent $\gamma_{dyn}$. This model describes
the data in Fig.\ \ref{fig:M/H}a over the entire temperature region above $T_{max}$.
A similar model can be formulated for the magnetization-field data in Fig.\
\ref{fig:M/H}b.

We define a crossover temperature $T_{cross}$ as the temperature where $\chi_1$ exceeds
$\chi_{dyn}$ (see Fig.\ \ref{fig:pd}). It can be regarded as the boundary of the
Griffiths phase. For $x=12.25\%$,
$T_{cross} \approx 7$K which is much higher than the cluster glass temperature 0.19K
found via the ac-susceptibility. This leaves room for a significant
superparamagnetic regime where independent frozen clusters dominate.
We note that superparamagnetism can arise in systems with
Ising spin symmetry even without rare region interactions \cite{millis02}. However,
our system does not show any indications of reduced spin symmetry.

Further analysis of the Curie term requires insight into the structure of the
rare region interactions. In a purely percolative scenario, one expects $A_1 \sim
|x-x_c|^{-\gamma_c}$ with the percolation exponent $\gamma_c=1.8$. Our data indeed
show a divergence of $A_1$ with $x \to x_c$ but with an exponent of about 2.6.
Alternatively, one can successfully model the Curie term as a contribution from a
number of frozen clusters of fixed moment which increases from $2\mu_B$ for $x=15\%$ to
$15 \mu_B$ for $x=12.07\%$ as $x_c$ is approached. A more detailed discussion will
be published elsewhere.

We emphasize that the theory \cite{vojta05,vojta0709} for quantum Griffiths effects in metals
was originally developed for antiferromagnets rather than ferromagnets. In ferromagnets,
mode-coupling effects produce an additional long-range interaction \cite{BKV} which renders
the disorder perturbatively irrelevant at the clean QCP \cite{NVBK99}.
However, this does not preclude the existence of Griffiths singularities because
they are nonperturbative degrees of freedom. In fact, it was recently shown
\cite{HoyosVojta07} that the physics of independent rare regions in a ferromagnet is
the same as in an antiferromagnet. We thus believe that the quantum Griffiths scenario
is applicable to our system, at least above the crossover temperature where
interactions between rare regions become important \cite{ferro_footnote}.

In summary, we have presented magnetization and susceptibility measurements of the transition
metal alloy  Ni$_{1-x}$V$_x$ close to the critical concentration for the onset of ferromagnetism.
While the finite-temperature phase transition in the concentrated Ni regime
($x\le11\%$) is well described by mean-field behavior, the diluted regime with low or vanishing $T_c$
cannot be described in terms of conventional critical behavior. Instead, the data follow the predictions
of a quantum Griffiths phase associated with an infinite-randomness
QCP over a wide temperature and field region.
Previous specific heat \cite{gregory75C} and transport data \cite{amamou75rho} support this scenario
via anomalous power laws in a wide concentration range (even though they were not discussed
in terms of a Griffiths phase).
Deviations at lower $T$ hint at individual freezing of large clusters before the system
enters a cluster glass phase.

We thank C. C. Almasan and S. D. Huang for the use of the SQUID magnetometer and the
X-ray diffractometer, respectively.
This work has been supported in part by the NSF under grant nos. DMR-0306766, DMR-0339147,
and DMR-0906566 as well as by Research Corporation


\begin{thebibliography}{19}
\bibitem{mag_footnote} According to the theory \cite{vojta0709},
 the typical moment should depend on temperature, but only very weakly
 (logarithmically) which is hard to confirm in any experiment.
\bibitem{ferro_footnote} This implies that a complete renormalization group theory of the itinerant
        ferromagnetic QPT would have to include the nonperturbative degrees of freedom.
\end{thebibliography}

\end{document}